\documentclass[pre,floatfix,twocolumn]{revtex4}
\usepackage{amsmath}
\usepackage{amsfonts}
\usepackage{mathrsfs}
\usepackage{epsfig}
\usepackage{graphicx}
\usepackage{dcolumn}
\usepackage{amssymb}
\usepackage{hyperref}
\usepackage[hyphenbreaks]{breakurl}

\begin{document}

\title{Deviations from universality in the fluctuation behavior of a
heterogeneous complex
system reveal intrinsic properties of components: The case of the
international currency market}

\author{Abhijit Chakraborty, Soumya Easwaran and Sitabhra
Sinha}

\affiliation{The Institute of Mathematical Sciences, CIT Campus,
Taramani, Chennai 600113, India.
}
\date{\today}
\begin{abstract}
Identifying behavior that is relatively invariant under different
conditions is a challenging task in far-from-equilibrium complex
systems. As an example of how the existence of a semi-invariant
signature can be masked by the heterogeneity in the properties of the
components comprising such systems, we consider the exchange rate
dynamics in the international currency market. We show that the
exponents characterizing the heavy tails of fluctuation distributions
for different currencies systematically diverge from a putative
universal form associated with the median value ($\simeq 2$) of the
exponents. We relate the degree of deviation of a particular currency
from such an ``inverse square law'' to fundamental macroscopic
properties of the corresponding economy, viz., measures of per capita
production output and diversity of export products. We also show that
in contrast to uncorrelated random walks exhibited by the exchange
rate dynamics for currencies belonging to developed economies, those
of the less developed economies show characteristics of sub-diffusive
processes which we relate to the anti-correlated nature of the
corresponding fluctuations. Approaches similar to that presented here
may help in identifying invariant features obscured by the
heterogeneous nature of components in other complex systems.
\end{abstract}

\maketitle

\section {Introduction}
The discovery that systems at equilibrium exhibit universality near a
phase transition has been a path-breaking achievement of statistical
physics in the previous century~\cite{Wilson1983}. However, despite considerable
effort, fluctuation behavior in biological and socio-economic systems
that are far from equilibrium are not yet well
understood~\cite{Henrickson2002}. Indeed,
strong evidence for universality of non-equilibrium transitions 
is still lacking~\cite{Hinrichsen2000}. The large diversity seen in
non-equilibrium critical phenomena 
poses a major challenge for those
trying to uncover general principles underlying the collective
dynamics of complex systems occurring in nature and society.
Such systems, apart from comprising a large number of
interacting components, are often characterized by a large degree of
heterogeneity in the properties of individual elements.
For example, components of a complex system may
exhibit qualitatively distinct dynamics. The local
connection density among the elements in
different parts may also greatly differ.
It is known that such heterogeneity can result in deviation from universal behavior expected near 
phase transitions~\cite{Cohen2002}.


A prototypical example of a complex system with a highly heterogeneous composition 
is the de-centralized international trade in foreign exchange (FOREX) which
constitutes the largest financial market in the world in terms of
volume~\cite{BIS2013}. 
An advantage of studying its fluctuation behavior over that of
other complex systems with many degrees of freedom is the
availability of large quantities of high-resolution digital data that
are relatively easily accessible for analysis~\cite{Yura2014}. 
The different currencies that are traded in the market are each subject to
multifarious influences, e.g., related to geographical, economic,
political or commercial factors, which can affect them in many
different ways. Such a highly heterogeneous system provides a stark
contrast to the relatively simpler systems having homogeneous
composition that have typically been investigated by physicists. 
In particular, we can ask whether the
components of a heterogeneous complex system can be expected to show
universal features, i.e., phenomena independent of microscopic
details, which may potentially be explained using tools of statistical
physics. For the specific case of the FOREX market, establishing any
robust empirical regularity will be an important contribution 
towards understanding the underlying self-organizing dynamics in
such systems. 
%
Note that the domain of microeconomics that is concerned with the
dynamics of single markets has seen accumulating
evidence suggestive of universality~\cite{Sinhabook}. 
The most robust of these relate to the nature
of the heavy-tailed distributions of fluctuations in individual stock
prices, as well as, equity market
indices~\cite{Jansen1991,Lux1996,Plerou1999,Pan2007}, often referred
to as the ``inverse cubic
law''~\cite{Gopikrishnan1998,Pan2008}. In contrast, 
macroeconomic processes have a
relative paucity of such ``stylized facts''.
Although the distribution of fluctuations in the
exchange rates of currencies has been the subject of several earlier
investigations~\cite{deVries1994,Friedrich2000}, some of which have indeed 
reported heavy tails for different currencies, there is little agreement concerning the
values of the power-law exponents characterizing such
tails - not even whether they lie outside the Levy-stable
regime~\cite{McFarland1982,Koedijk1990,Loretan1994,Guillaume1997}.
This suggests that the nature of the fluctuation distribution for a
particular
currency could be related to some intrinsic properties of the
underlying economy.

In this paper we show that there is indeed a systematic deviation from
a putative universal signature - which we refer to as ``inverse square
law'' - for the fluctuation behavior of different currencies depending
on two key macroeconomic indicators, viz., the gross domestic product
(GDP) per capita related
to the economic
performance, and the Theil index that measures the diversity of
exports of the corresponding
countries (see data description for details). Thus, several underdeveloped (frontier) economies exhibit currency
fluctuations whose distributions appear to be of a Levy-stable nature,
while those of most developed economies fall outside this regime.
The median value of the exponents quantifying the heavy-tailed
nature of the cumulative fluctuation distributions for all the currencies occur
close to 2, i.e., at the boundary of the Levy-stable regime.
Our study demonstrates how robust empirical
regularities in complex systems can be uncovered when they are masked by the intrinsic
heterogeneity among the individual components. 
We have also characterized the distinct nature of the exchange rate dynamics of different
currencies by considering their self-similar
scaling behavior. Our analysis reveals that while currencies of
developed economies follow uncorrelated random walks, those of
emerging and frontier economies exhibit sub-diffusive
(or mean-reverting) dynamics. 

\section{Data description}
The data-set we have analyzed comprises the daily
exchange rates with respect to the US Dollar (USD) of $N=75$
currencies (see Table~\ref{table1})
for the period October 23, 1995 to April 30, 2012,
corresponding to $\tau=6035$ days.
The rate we use is the midpoint value, i.e., the average of the bid and ask
rates for 1 USD against a given currency. The data is obtained from a
publicly accessible archive of historical interbank market rates
maintained by the
Oanda corporation, an online currency conversion site~\cite{Oanda}
that is used by major corporations, tax authorities and auditing
firms worldwide.
The interbank (or spot) rate for a currency is
the official rate quoted in the media and that apply
to large transactions of $10^6$ USD or higher (typically taking place
between banks and financial institutions).
For each day, the site records an average value that is calculated
over all rates collected over a 24 hour
period from frequently updated sources in
the global foreign exchange market,
including online currency trading platforms, leading market data
vendors, and contributing financial institutions.
We have chosen USD as the base currency for the exchange rate as it is
the preferred currency for most international transactions and
remains the reserve currency of choice for most
economies~\cite{Papaioannou2006,IMFAnnRep}.
We have verified that using other base currencies lead to
qualitatively similar fluctuation distributions for exchange rates.

\begingroup
\squeezetable
\begin{table*}
\caption{The currencies of developed (1-14), emerging (15-44) and
frontier (45-75) economies considered in the study. The columns
indicate the currency code along with the nature of the exchange rate
regime (as obtained from Oanda and XE sites), the character of the
economy (as categorized by MSCI), the geographical region, the average
GDP per capita (provided by IMF) and the mean Theil index (calculated
from data available from MIT OEC) for the corresponding countries.}
\begin{ruledtabular}
\begin{tabular}{c c c c c c c c }
Sl. no.  &  Currency                &  Code   & Exchange Rate Regime    &  Market Type       &  Region      &  $\langle g \rangle$ in USD     &   $\langle T \rangle$ \\
 &  &  &  (Oanda, XE) & (MSCI)  &  & (IMF)  &  (MIT) \\ \hline \hline
1 &  Canadian Dollar                &  CAD    & Floating              &  Developed         & Americas     &  32561.46       &  1.95 \\
2 &  Danish Krone                   &  DKK    & Pegged within horizontal band         &  Developed         & Europe       & 44617.1         & 1.49 \\
3 &  Euro                           &  EUR    & Floating              &  Developed         & Europe       & 28200.99        & - \\
4 &  Great Britain Pound            &  GBP    & Floating              &  Developed         & Europe       & 32126.2         & 1.54 \\
5 &  Iceland Krona                  &  ISK    & Floating              &  Developed         & Europe       & 39213.54        & 3.69 \\
6 &  Norwegian Kroner               &  NOK    & Floating              &  Developed         & Europe       & 59286.29        & 3.45 \\
7 &  Swedish Krona                  &  SEK    & Floating              &  Developed         & Europe       & 39571.51        & 1.63 \\
8 &  Swiss Franc                    &  CHF    & Floating              &  Developed         & Europe       & 52059.39        & 1.96 \\
9 &  Israeli New Shekel             &  ILS    & Floating              &  Developed         & Middle East  & 22478.26        & 2.64 \\
10 &  Australian Dollar             &  AUD    & Floating              &  Developed         & Asia-Pacific & 35251.16        & 2.38 \\
11 &  Hong Kong Dollar              &  HKD    & Fixed peg         &  Developed         & Asia-Pacific & 27406.74        & 1.98 \\
12 &  Japanese Yen                  &  JPY    & Floating              &  Developed         & Asia-Pacific & 36942.47        & 1.95 \\
13 &  New Zealand Dollar            &  NZD    & Floating              &  Developed         & Asia-Pacific & 23459.35        & 2.14 \\
14 &  Singapore Dollar              &  SGD    & Floating              &  Developed         & Asia-Pacific & 30538.39        & 2.65 \\
\hline
15 &  Bolivian Boliviano            &  BOB    & Crawling peg              &  Emerging          & Americas     & 1287.16         & 3.65 \\
16 &  Brazilian Real                &  BRL    & Floating              &  Emerging          & Americas     & 6254.18         & 1.93 \\
17 &  Chilean Peso                  &  CLP    & Floating              &  Emerging          & Americas     & 7563.51         & 3.23 \\
18 &  Colombian Peso                &  COP    & Floating              &  Emerging          & Americas     & 3864.52         & 3.01 \\
19 &  Dominican Republic Peso       &  DOP    & Floating              &  Emerging          & Americas     & 3509.27         & 2.84 \\
20 &  Mexican Peso                  &  MXN    & Floating              &  Emerging          & Americas     & 7556.32         & 2.15 \\
21 &  Peruvian Nuevo Sol            &  PEN    & Floating              &  Emerging          & Americas     & 3243.22         & 2.99 \\
22 &  Venezuelan Bolivar            &  VEB    & Fixed peg         &  Emerging          & Americas     & 6302.1          & 4.85 \\
23 &  Albanian Lek                  &  ALL    & Floating              &  Emerging          & Europe       & 2319.21         & 2.77 \\
24 &  Czech Koruna                  &  CZK    & Floating              &  Emerging          & Europe       & 11701.17        & 1.44 \\
25 &  Hungarian Forint              &  HUF    & Pegged within horizontal band              &  Emerging          & Europe       & 9151.13         & 1.87 \\
26 &  Polish Zloty                  &  PLN    & Floating              &  Emerging          & Europe       & 7866.73         & 1.41 \\
27 &  Russian Rouble                &  RUB    & Floating              &  Emerging          & Europe       & 5791.06         & 3.23 \\
28 &  Turkish Lira                  &  TRY    & Floating              &  Emerging          & Europe       & 6451.81         & 1.58 \\
29 &  Algerian Dinar                &  DZD    & Floating              &  Emerging          & Africa       & 2890.28         & 5.17 \\
30 &  Cape Verde Escudo             &  CVE    & Fixed peg         &  Emerging          & Africa       & 2130.76         & 3.71 \\
31 &  Egyptian Pound                &  EGP    & Floating              &  Emerging          & Africa       & 1727.67         & 2.73 \\
32 &  Ethiopian Birr                &  ETB    & Floating              &  Emerging          & Africa       & 208.91          & 4.33 \\
33 &  Mauritius Rupee               &  MUR    & Floating              &  Emerging          & Africa       & 5432.83         & 3.39 \\
34 &  Moroccan Dirham               &  MAD    & Fixed peg              &  Emerging          & Africa       & 1997.64         & 2.54 \\
35 &  South African Rand            &  ZAR    & Floating              &  Emerging          & Africa       & 4751.66         & 2.14 \\
36 &  Tanzanian Shilling            &  TZS    & Floating              &  Emerging          & Africa       & 361.18          & 3.17 \\
37 &  Chinese Yuan Renminbi         &  CNY    & Fixed peg             &  Emerging          & Asia         & 2173.96         & 1.55 \\
38 &  Indian Rupee                  &  INR    & Floating              &  Emerging          & Asia         & 774.57          & 1.74 \\
39 &  Indonesian Rupiah             &  IDR    & Floating              &  Emerging          & Asia         & 1630.89         & 1.99 \\
40 &  Papua New Guinea Kina         & PGK  & Floating                 &  Emerging          & Asia         & 1014.91         & 4.34 \\
41 &  Philippine Peso               &  PHP    & Floating              &  Emerging          & Asia         & 1440.56         & 3.05 \\
42 &  South Korean Won              &  KRW    & Floating              &  Emerging          & Asia         & 15655           & 2.11 \\
43 &  Taiwan Dollar                 &  TWD    & Floating              &  Emerging          & Asia         & 15707.7         & - \\
44 &  Thai Baht                     &  THB    & Floating              &  Emerging          & Asia         & 3194.12         & 1.78 \\
\hline
45 &  Guatemalan Quetzal            &  GTQ    & Floating              &  Frontier          & Americas     & 2134.53         & 2.54 \\
46 &  Honduran Lempira              &  HNL    & Crawling peg              &  Frontier          & Americas     & 1380.32         & 3.23 \\
47 &  Jamaican Dollar               &  JMD    & Floating              &  Frontier          & Americas     & 4042.29         & 4.25 \\
48 &  Paraguay Guarani              &  PYG    & Floating              &  Frontier          & Americas     & 1892.51         & 3.81 \\
49 &  Trinidad Tobago Dollar        &  TTD    & Floating              &  Frontier          & Americas     & 12983.73        & 4.21 \\
50 &  Croatian Kuna                 &  HRK    & Floating              &  Frontier          & Europe       & 9166.72         & 1.75 \\
51 &  Kazakhstan Tenge              &  KZT    & Floating              &  Frontier          & Europe       & 4399.18         & 3.97 \\
52 &  Latvian Lats                  &  LVL    & Fixed peg         &  Frontier          & Europe       & 6912.26         & 2.35 \\
53 &  Botswana Pula                 &  BWP    & Crawling peg              &  Frontier          & Africa       & 5447.23         & 5.45 \\
54 &  Comoros Franc                 &  KMF    & Fixed peg         &  Frontier          & Africa       & 609.64          & 5.05 \\
55 &  Gambian Dalasi                &  GMD    & Floating              &  Frontier          & Africa       & 513.84          & 4.27 \\
56 &  Ghanaian Cedi                 &  GHC    & Floating              &  Frontier          & Africa       & 871.99          & 4.11 \\
57 &  Guinea Franc                  &  GNF    & Fixed peg              &  Frontier          & Africa       & 419.31          & 5.03 \\
58 &  Kenyan Shilling               &  KES    & Floating              &  Frontier          & Africa       & 586             & 2.95 \\
59 &  Malawi Kwacha                 &  MWK    & Floating              &  Frontier          & Africa       & 226.69          & 4.5 \\
60 &  Mauritanian Ouguiya           &  MRO    & Floating              &  Frontier          & Africa       & 752.88          & 4.96 \\
61 &  Mozambique Metical            &  MZM    & Floating              &  Frontier          & Africa       & 330.54          & 4.28 \\
62 &  Nigerian Naira                &  NGN    & Floating              &  Frontier          & Africa       & 782.49          & 6.02 \\
63 &  Sao Tome and Principe Dobra   &  STD    & Fixed peg         &  Frontier          & Africa       & 864.42          & 4.15 \\
64 &  Zambian Kwacha                &  ZMK    & Floating              &  Frontier          & Africa       & 677.53          & 4.67 \\
65 &  Jordanian Dinar               &  JOD    & Fixed peg         &  Frontier          & Middle East  & 2617.85         & 3 \\
66 &  Kuwaiti Dinar                 &  KWD    & Fixed peg              &  Frontier          & Middle East  & 25554.56        & 5.49 \\
67 &  Syrian Pound                  &  SYP    & Fixed peg             &  Frontier          & Middle East  & 1732.47         & 4.21 \\
68 &  Brunei Dollar                 &  BND    & Fixed peg  &  Frontier          & Asia         & 23516.1         & 5.45 \\
69 &  Bangladeshi Taka              &  BDT    & Floating              &  Frontier          & Asia         & 436.09          & 3.63 \\
70 &  Cambodian Riel                &  KHR    & Floating              &  Frontier          & Asia         & 498.85          & 3.85 \\
71 &  Fiji Dollar                   &  FJD    & Fixed peg              &  Frontier          & Asia         & 3052.58         & 3.37 \\
72 &  Lao Kip                       &  LAK    & Floating              &  Frontier          & Asia         & 577.82          & 3.66 \\
73 &  Pakistan Rupee                &  PKR    & Floating              &  Frontier          & Asia         & 756.7           & 2.87 \\
74 &  Samoan Tala                   &  WST    & Fixed peg              &  Frontier          & Asia         & 2215.73         & 4.63 \\
75 &  Sri Lankan Rupee              &  LKR    & Floating   &  Frontier          & Asia         & 1438.09         & 2.65 \\
\end{tabular}
\end{ruledtabular}
\label{table1}
\end{table*}
\endgroup


The choice of currencies used in our study is mainly dictated by the
exchange rate regime (see Table~\ref{table1}), which is
obtained from the site~\cite{Oanda} where we collected the exchange
rates data and supplemented by information from the site of another
online FOREX services company~\cite{xe}.
In particular, we have not considered currencies
whose exchange rate with respect to USD is constant over time.
Most of the currencies in our database are floating, either freely
under the influence of market forces or managed to an extent with no
pre-determined path. Among the remaining currencies, a few are pegged
to USD or some other important currency (such as EUR), but with
some variation within a band (which may either be fixed or moving in
time). Note that as the EUR was introduced in January 1, 1999, i.e.,
within the time interval considered by us, we have used the exchange
rate for the ECU (European Currency Unit) for the period October 23,
1995 to December 31, 1998.

To ensure that the observed differences in the nature of the
fluctuation distributions of currencies is not just a trivial outcome
of the different exchange rate regimes, we have performed a two-sample
Kolmogorov-Smirnov test~\cite{Marsaglia2003} with the null hypothesis
that the pegged
and floating currencies are sampled from the same continuous
distribution. A measured $p$-value of 0.39 indicates that the null
hypothesis cannot be rejected at $5\%$ level of
significance. We also carried out a Wilcoxon
rank sum test~\cite{Gibbons2011}
with the null hypothesis that both pegged
and floating currencies are sampled from continuous distributions with
equal medians. We obtained a $p$-value of 0.27, again
indicating that there is not
enough evidence to reject the null hypothesis  at $5\%$ level of
significance. We thus conclude that the distinct behavior of the
currencies in terms of the distribution of their exchange rate returns
cannot be simply explained away as being related to their pegged
or floating nature.

In order to explore whether the nature of the fluctuation distribution of a
particular currency could be related to the characteristics of the
underlying economy, the countries to which these
currencies belong are grouped into three categories,
viz., developed, emerging and frontier markets, as per the Morgan
Stanley Capital International (MSCI) market classification
framework~\cite{MSCI}. This is done on the basis of several criteria
such as, the sustainability of economic development,
number of companies meeting certain size and liquidity criteria, ease
of capital flow, as
well as, efficiency and stability of the institutional framework.

To make more explicit the connection between deviation from universality and the
heterogeneity of the constituents of the FOREX market, we have examined in
detail certain macro-economic factors characterizing a national
economy for the role they may play in
determining the nature of
the fluctuation dynamics of a currency. In particular, we find that a
prominent role is played by
(a) the GDP per capita $g$, as well as, (b) the Theil
index $T$ of export products, which we define below.

The {\em GDP per capita} of a country is obtained by dividing the
annual economic output, i.e., the aggregate value
of all final goods and services produced in it during a year,
by the total population. It is one of the primary
indicators of the economic performance of a country, with higher GDP
per capita indicating a higher standard of living for the people
living in it~\cite{GDP}.
The annual GDP per capita of the countries whose currencies have
been included in our study are obtained from publicly accessible data
available in the website of
the International Monetary Fund (IMF)~\cite{IMF}. We have averaged the
data over the 18 year period (1995-2012) considered in our study to
obtain the mean GDP per capita $\langle g \rangle$.

The {\em Theil index} measures the diversity of the export products of a
country~\cite{Theil1967} and is defined as
$T = \frac{1}{M}\sum_{i=1}^M ( \frac{x_i}{\bar{x}}
\ln{\frac{x_i}{\bar{x}}}),$
where $x_i$ is the total value (in USD) of the $i$-th export product
of a country,
${\bar{x}}$ is the average value of all export products and $M$ is
the total number of different products that are exported.
A high value of $T$ corresponds to large heterogeneity in the values of the
different exported products, indicating that a few products dominate
the export trade. By contrast, low $T$ implies that a country has a
highly diversified portfolio of export products and therefore,
relatively protected from the vagaries of fluctuations in the demand for any
single product.
To compute the Theil index we have used
the annual export product data of different countries available from
the Observatory
of Economic Complexity (OEC) at MIT~\cite{mit}. We have used the four digit
level of the Standard International Trade Classification for
categorizing different products which corresponds to $M = 777$ distinct
export products in the data set. We have averaged the annual Theil
indices over the period 1995-2012 to obtain the mean Theil index
$\langle T \rangle$ for each
country.

\begin{figure}[t]
\begin{center}
\includegraphics[width=0.99\linewidth,clip]{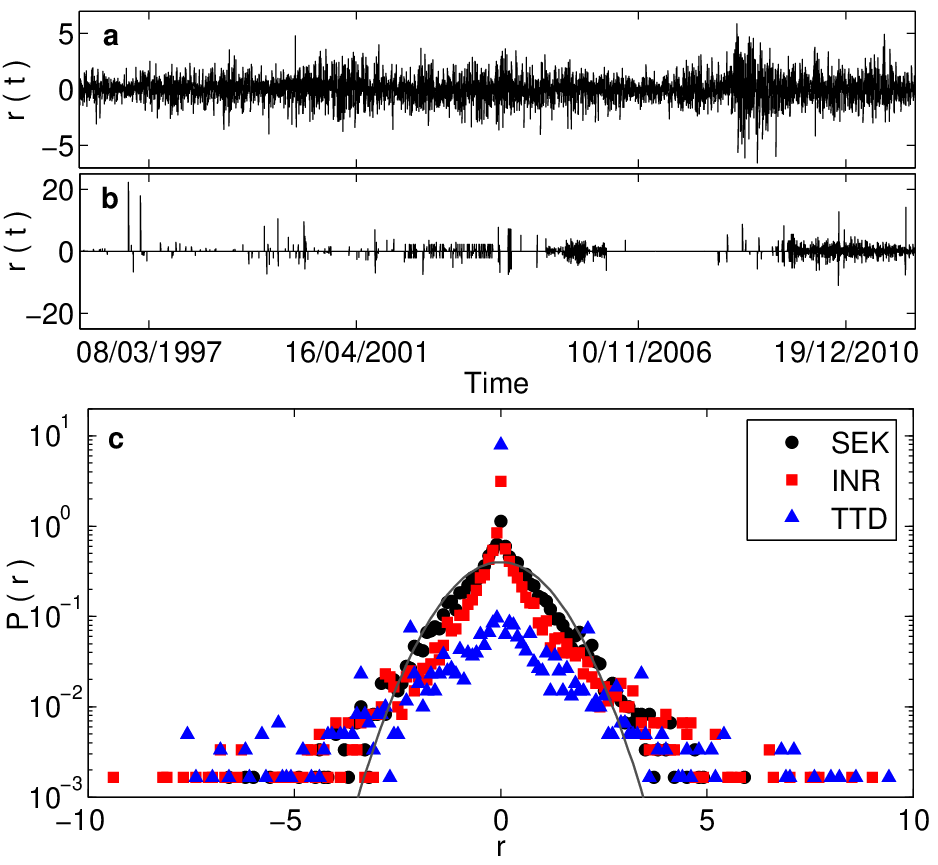} 
\end{center}
\caption{(color online). 
{\bf Heavy tailed behavior in the distribution of currency 
exchange rate fluctuations.} The time-series of normalized log returns
$r(t)$ for currencies of developed economies, e.g., SEK (a), shows
relatively lower amplitude variations compared to that of
currencies of frontier economies, e.g., TTD (b), in general (note the
different scales in the ordinate of the two panels). However, the
probability density functions of $r$ for all currencies show a heavy-tailed nature,
shown in (c) for currencies from a developed economy, SEK (black, circles), an
emerging economy, INR (red, squares), and a frontier economy, TTD
(blue, triangles). For comparison, the standard normal distribution is shown
using a solid curve.
}
\label{fig1}
\end{figure}
\begin{figure}[!h]
\begin{center}
\includegraphics[width=0.99\linewidth,clip]{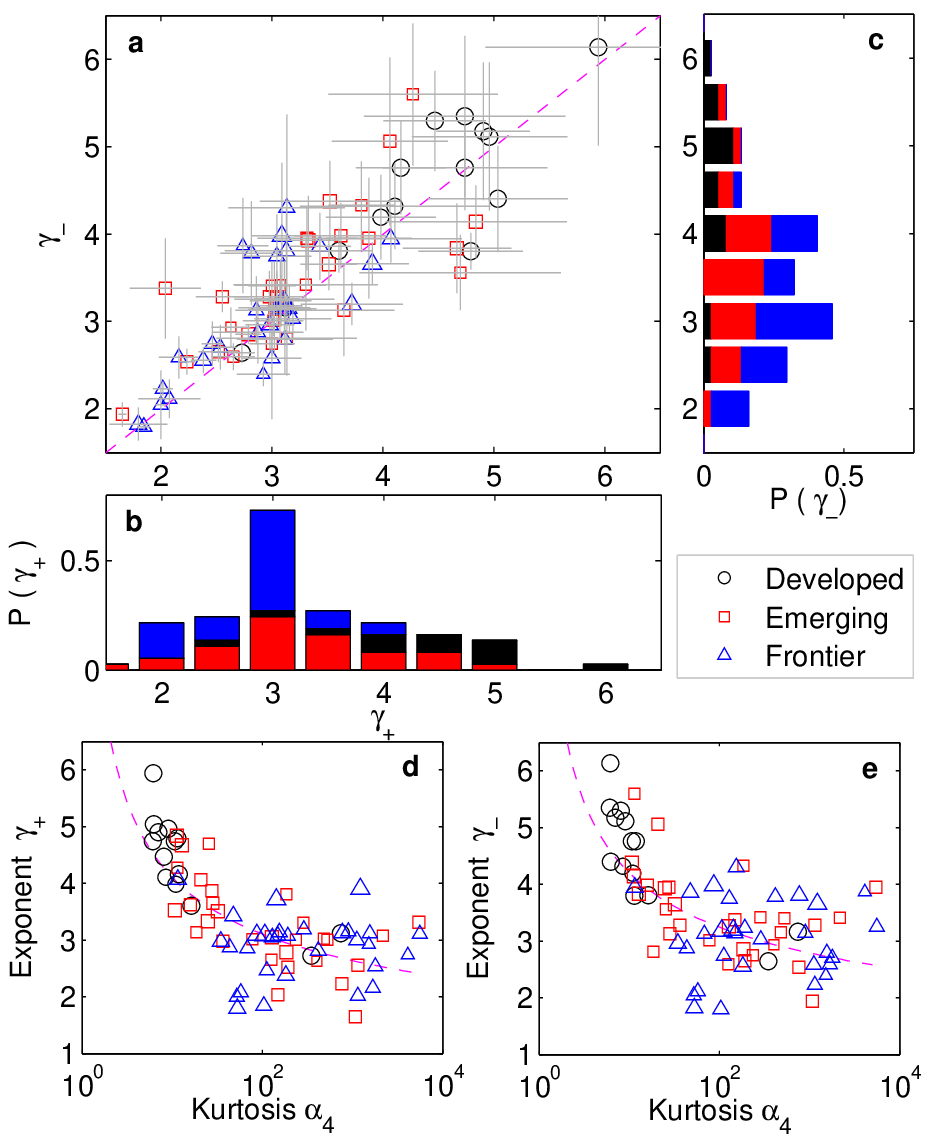} 
\end{center}
\caption{(color online).
{\bf (a-c) Deviation from universality for
exchange rate fluctuations.} The probability distribution of the 
power law exponents $\gamma_+$ (b) and $\gamma_-$ (c) obtained by maximum 
likelihood estimation (MLE)
for the positive and negative tails, respectively, of the individual
return distributions for the 75 currencies, show a peak around 3 with
median values of 3.11 (for $\gamma_+$) and 3.28 (for $\gamma_-$). 
Error bars indicate the uncertainty in the estimated values
and are obtained by a non-parametric bootstrap technique. 
Points lying
closer to the diagonal ($\gamma_+ = \gamma_-$, indicated by a broken
line) in (a) imply a higher degree of symmetry in the distribution of
$r$ for the corresponding currency, i.e., positive and negative fluctuations of similar
magnitude are equally probable. 
The heavy-tailed nature of the distributions characterized by the
tail-exponents correspond closely to
their peakedness measured using the kurtosis $\alpha_4$, as shown
by the scatter plot between (d) $\alpha_4$ and $\gamma_+$ and (e)
$\alpha_4$ and $\gamma_-$ for the currencies. The best log-linear
fits, indicated by broken lines, correspond to $\alpha_4 = {\rm exp}[(\gamma_{\pm}/A_{\pm})^{-\beta_{\pm}}]$ with
$A_+ = 5.8$, $\beta_+ = 2.4$ (d) and $A_-= 5.6$, $\beta_- = 2.8$ (e).
The Pearson correlation
coefficient between log(log($\alpha_4$)) and log($\gamma_{\pm}$) are
$\rho = -0.67$ ($p = 10^{-11}$) for (d) and $\rho = -0.59$
($p = 10^{-8}$) for (e).
Different symbols and colors are used to indicate
currencies from developed (black, circles), emerging (red, squares) and 
frontier (blue, triangles) economies, while symbol size is 
proportional to log($\langle g \rangle$) of the corresponding countries.
}
\label{fig2}
\end{figure}
\section{Results}
We have measured the fluctuations in the exchange rates of
75 currencies (see data description for details) with respect to the US Dollar
over the period 1995-2012. To ensure that the result is
independent of the unit of measurement,
we have quantified the variation in the exchange
rate $P_i(t)$ of the $i$-th currency ($i = 1, \ldots, N$) at time $t$
by its logarithmic return defined over a time-interval $\Delta t$ as 
$R_i(t,\Delta t) = ln \, P_i(t+\Delta t)-ln\, P_i(t) $. 
As explained in the data description, our data comprises daily
exchange rates and we therefore consider $\Delta t = 1$ day.
Different currencies can
vary in terms of the intensity of fluctuations in their exchange
rates (volatility) as can be measured
by the standard deviation $\sigma$ of the returns. Thus, to compare
the return distributions of the different currencies, we normalize the
returns of each currency $i$ by subtracting the mean value $\langle
R_i \rangle = \Sigma_{t=1}^{\tau-1} R_i (t)/(\tau-1)$ and
dividing by the standard deviation $\sigma_i (t) =\sqrt{\frac{1}{\tau-2}
\Sigma_{t^\prime \neq t} [R_i (t^\prime) - \langle R_i \rangle]^2}$
(removing the self contribution from the measure of 
volatility), obtaining
the normalized return, $ r_i(t)=(R_i(t)-\langle R_i
\rangle)/\sigma_i(t)$.

\subsection{The ``inverse square law" of the distribution of
fluctuations for currency exchange rates}
As can be seen from Fig.~\ref{fig1}~(a-b), the returns quantifying the
fluctuations in the exchange rate of currencies can appear extremely
different even though they have been normalized by their volatilities.
The temporal variation of $r(t)$ for SEK [shown in
Fig.~\ref{fig1}~(a)], the currency of a developed
economy, is mostly bounded between a narrow interval around 0 with the
fluctuations never exceeding 6 standard deviations from the mean
value. By contrast, Fig.~\ref{fig1}~(b) shows that TTD, belonging to a
frontier economy, frequently exhibits extremely large
fluctuations that can occasionally exceed even 20 standard deviations - 
an event
extremely unlikely to have been observed had the distribution been of a
Gaussian nature~\cite{note}.
These observations suggest that the distributions of the exchange rate
fluctuations have long tails and that different currencies may have
significantly different nature of heavy-tailed behavior.
As shown in Fig.~\ref{fig1}~(c), where the distributions of $r$ for
SEK, TTD and an emerging economy currency, INR, is displayed, this is
indeed the case. 

The nature of the tails of the return distributions is
established quantitatively by fitting them to a power-law decay 
for the probability distribution having the 
functional form $P(r) \sim r^{-\gamma}$ through maximum
likelihood estimation (MLE)~\cite{Clauset}. Uncertainty in estimating the
optimal value of $\gamma$ is calculated by performing MLE of exponents from
100 surrogate data-sets for each currency. These are
constructed by random sampling with replacement from the original return
time-series data~\cite{Clauset}.
While both the positive and negative returns show heavy tails, we note
that the exponents characterizing them need not be identical
for a currency, such that the corresponding return distribution is
asymmetric or skewed. The scatter plot in Fig.~\ref{fig2}~(a) shows how the
positive and negative tail exponents, $\gamma_+$ and $\gamma_-$
respectively, are related to each other for the different currencies.
Currencies that occur closer to the diagonal line $\gamma_+=\gamma_-$
have similar nature of upward and downward exchange rate movements.
However, currencies which occur much above the diagonal  (i.e.,
$\gamma_+ < \gamma_-$) will tend to have a higher probability of
extreme positive returns compared to negative ones, while those below
the diagonal are more likely to exhibit very large negative
returns. 
We note in passing that the skewness depends, to some extent, on the state of the
economy of the country to which a currency belongs, with return
distributions of developed economies being the least asymmetric in
general, having mean skewness $0.52 \pm
1.28$, while those of emerging and frontier economies are relatively
much higher, being $6.54 \pm
15.24$ and $6.60 \pm 18.04$, respectively.
 
The distribution of the exponents
characterizing the power-law nature of the exchange-rate returns shown in
Figs.~\ref{fig2}~(b-c) peaks around 3 for both the positive and
negative tails. As a probability distribution function with a power
law characterized by exponent value $\gamma \simeq 3$ implies that the
corresponding CCDF also has a power-law form but with exponent value
$\alpha = \gamma - 1 \simeq 2$~\cite{Newman2005}, this result suggests an ``inverse
square law'' governing the
nature of fluctuations in the currency market in contrast to the 
``inverse cubic law'' that has been proposed as governing the price
and index fluctuations in several financial
markets~\cite{Jansen1991,Lux1996,Gopikrishnan1998,Plerou1999,Pan2007,Pan2008}.
However, as is the case here, such a ``law'' is only manifested on the average, as
the return distributions for individual assets can have quite distinct 
exponents~\cite{Pan2007}. Here, we observe that the different
currencies can have exponents as low as 2 and as high as 6. Moreover,
there appears to be a strong correlation between the nature of the
tail and the state of the underlying economy to which the currency
belongs. Thus, developed economy currencies tend to have the largest
exponents, while most of the lowest values of exponents belong to
currencies from the frontier economies. 
This suggests an
intriguing relation between the nature of currency fluctuations and the state of the
underlying economy, that could possibly be quantified by one or more 
macroeconomic indicators. This theme is explored in detail below.
\begin{figure}
\begin{center}
\includegraphics[width=0.99\linewidth,clip]{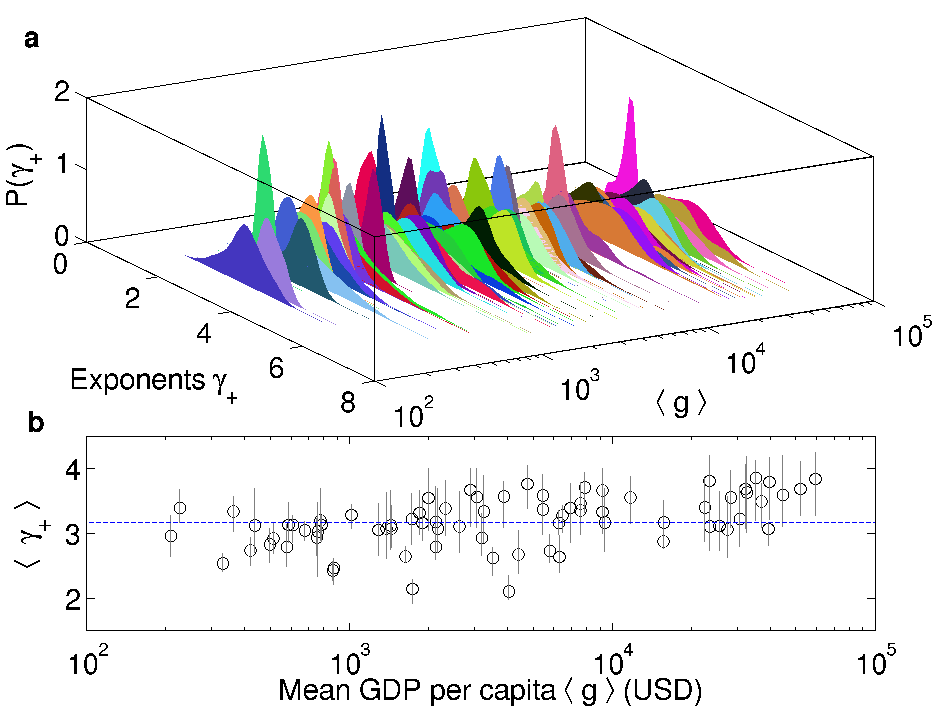} 
\end{center}
\caption{(color online).
{\bf Distribution of fluctuations for currency exchange
rates with respect to different base currencies display an ``inverse
square law'' on average.}
(a) The ensemble of distributions of power-law exponents $\gamma_{+}$ for the
positive return distributions of 75 currencies calculated with respect
to each of 75 base currencies that are arranged according to the mean 
GDP per capita $\langle g \rangle$ of the corresponding economy (of
the base currency).
(b) The mean values of the exponents $\gamma_{+}$ (circles) obtained using each
of the base currencies are almost all clustered around the value of 3,
indicating an ``inverse square law'' behavior of the heavy tails of
return distributions that is relatively stable against the
choice of different bases for measuring the fluctuation.
Error bars indicated represent the standard deviation in the estimated
values of $\gamma_{+}$ for different currencies for a given base
currency.
The broken line represents the grand average ($\langle \langle \gamma_{+} \rangle
\rangle = 3.21$) of the values for the
the exponent $\gamma_{+}$, taken over all currencies and bases.
}
\label{fig2B}
\end{figure}

In order to verify that the nature of fluctuations in exchange rates
does not change drastically depending on the specific choice of base
currency, we have re-calculated the exponents $\gamma$
characterizing return distributions of different currencies that are obtained
using each of the 75 currencies as the base. Fig.~\ref{fig2B} shows that for
all base currencies used in our study, the exponents $\gamma_+$ are 
distributed
about mean values that fluctuate around $\langle \langle \gamma_{+}
\rangle \rangle \sim 3$ (a similar behavior is seen for the exponents
of the negative returns distributions, $\gamma_-$).
This suggests that the “inverse square law” form for the
heavy tails of return distributions is valid {\em on average} relatively
independent of the base currency used to calculate the exchange rates.

The character of the heavy tails of the returns $r$ is closely related to the
peaked nature of the distribution that can be quantified
by its kurtosis which is defined as $\alpha_4 = E(r-\mu)^4/\sigma^4$,
where $E( )$ is the expectation while $\mu$ and $\sigma$ are the mean
and standard deviation, respectively, of $r$. 
Fig.~\ref{fig2}~(d-e) shows the relation between the kurtosis and the
exponents for the tails of the return distributions of the different
currencies. The fitted curve shown qualitatively follows the
theoretical relation between the two which can be derived by assuming
that the distribution is Pareto, i.e., follows a power law (although for such a situation, the
kurtosis is finite only for exponent values $\gamma >5$). 
We observe that the relation between the exponents and kurtosis
suggested by the scatter plots can be
approximately fit by the function $\alpha_4 \sim {\rm
exp}[(\gamma_{\pm}/A_{\pm})^{-\beta_{\pm}}]$ with
$\beta_+ = 2.4$, $A_+ = 5.8$ for the positive tail and 
$\beta_- = 2.8$, $A_- = 5.6$ for the negative tail
[Fig.~\ref{fig2}~(d) and (e),respectively]. The strong 
correlation between the peakedness of the distribution and the
character of the heavy tails can be quantified by the
Pearson correlation coefficients between log($\gamma_{\pm}$) and
log(log($\alpha_4$)), viz., $\rho = -0.67$ ($p = 10^{-11}$) for
the positive returns and $\rho = -0.59$ ($p = 10^{-8}$) for the
negative returns. Thus, instead of using two
different exponent values (corresponding to the positive and negative
tails) for each return distribution, we shall henceforth focus on the 
single kurtosis value that characterizes the distribution.

\subsection{Deviation from universality related to macroeconomic
factors}

\begin{figure}
\begin{center}
\includegraphics[width=0.99\linewidth,clip]{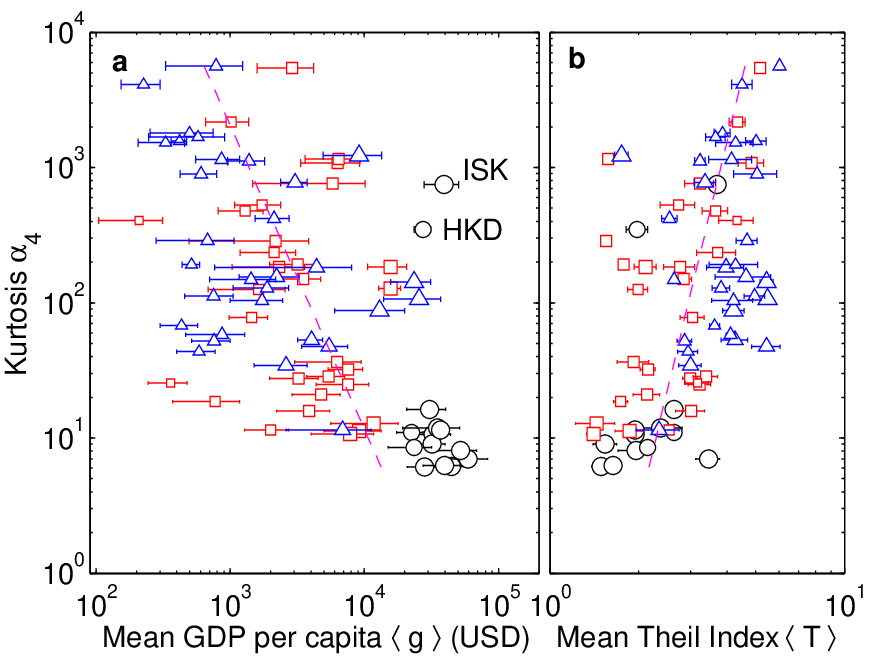} 
\end{center}
\caption{(color online). 
{\bf Variation of the kurtosis $\alpha_4$ of exchange rate fluctuation
distributions of different currencies with (a) annual GDP per capita,
$\langle g \rangle$ (in USD)
and (b) annual Theil index of the export products, $\langle T \rangle$, for the corresponding
countries, averaged over the period 1995-2012.}
The Pearson correlation coefficient between log($\langle g \rangle$) and
log($\alpha_4$) is $\rho = -0.55$ ($p = 10^{-7}$), the best-fit functional relation 
between the two being $\alpha_4 \sim \langle g \rangle^{-2.2}$.
Currencies of developed economies that are outliers from
this general trend, viz., ISK and HKD that have
high kurtosis despite having high GDP per capita,
are explicitly indicated in (a).
A similar analysis shows that the Pearson correlation coefficient between
log($\langle T \rangle$) and log($\alpha_4$) is $\rho = 0.53$ ($p = 10^{-6}$), with the
best-fit functional relation being $\alpha_4 \sim \langle T \rangle^{9.1}$.
Different symbols are used to indicate currencies from developed
(black, circles), 
emerging (red, squares) and frontier (blue, triangles) economies, while symbol
size is proportional to
log($\langle g \rangle$) of the corresponding countries.
Error bars represent the standard deviation of the annual values of $g$ and
$T$ over the period 1995-2012 for the countries corresponding to each
currency. 
}
\label{fig3}
\end{figure}

Given the variation in the nature of fluctuation distribution of
different currencies from a single universal form, we ask whether the
deviations are systematic in nature. Note that, the currencies 
belong to countries having very diverse economies, that trade in
a variety of products \& services with other countries and which may have
contrasting economic performances. 
An intuitive approach would be to relate the differences in the return
distributions with metrics which capture important aspects of the
economies as a whole. Fig.~\ref{fig3} shows that there is indeed a
significant correlation between the kurtosis of the return
distributions for the currencies and two macroeconomic indicators of
the underlying economies, viz., the mean GDP per capita, $\langle g
\rangle$, and the mean Theil
index, $\langle T \rangle$, that describe the overall prosperity and
the diversity of export products, respectively (see data description for details). 

Fig.~\ref{fig3}~(a) shows that the scatter of kurtosis $\alpha_4$
against $\langle g \rangle$ can be approximately fit by a power law of the form: 
$\alpha_4 \sim \langle g\rangle^{-2.2}$. The Pearson correlation coefficient between
the logarithms of the two quantities is $\rho = -0.55$ ($p = 10^{-7})$. 
Thus, in general, currencies of countries having higher GDP per capita
tend to be more stable, in the sense of having low probability of
extremely large fluctuations. However, there are exceptions
where currencies
exhibit high kurtosis even when they belong to countries with high GDP
per capita (e.g., HKD and ISK which are indicated in the figure). 
In these cases, the peakedness of the distribution
may reflect underlying economic crises, e.g., the 2008 Icelandic
financial crisis in the case of ISK and the 2003 SARS crisis for HKD.
Furthermore, we observe that currencies belonging to high GDP per capita
economies that are dependent on international trade of a few key
resources - such as, crude oil - also exhibit high kurtosis 
(e.g., KWD and BND).
This suggests a dependence of the nature of the fluctuation
distribution on the diversity of their exports, which is indeed shown
in Fig.~\ref{fig3}~(b). The dependence of the kurtosis on
$T$ (which is a measure of the variegated nature of trade) of the
corresponding economy is approximately described by a power-law
relation: $\alpha_4 \sim \langle T \rangle^{9.1}$. The Pearson correlation coefficient
between the logarithms of the two quantities is $\rho =0.53$ ($p =
10^{-6})$. This implies that, in general, currencies of countries
having low $\langle T \rangle$, i.e., having well-diversified export profile, tend to
be more stable.

Note that the fluctuations of the currencies depend on both of these
above macroeconomic factors, and the differences in their nature
cannot be explained exclusively by any one of them. It is therefore
meaningful to perform a multi-linear regression of $\alpha_4$ as a
function of both GDP per capita and Theil index using an equation of 
the form: log($\alpha_4$) $= b_0 + b_1$ log($\langle g \rangle$)$ + b_2$ log($\langle T \rangle$), where 
the constants $b_0 (=6.74), b_1 (=-0.48)$ and $b_2 (=1.69)$ are the best-fit
regression coefficients.
The coefficient of determination $R^2$, which measures how well the
data fits the statistical model, is found to be $0.39$ ($p
\simeq10^{-8}$).
This indicates that together the macroeconomic factors of GDP per
capita (related to the overall economic performance) and Theil index
(related to the international trade of the country) explain over
$39\%$ of the variation between the nature of the return distributions
of the different currencies.

One of the assumptions of multi-linear regression analysis is that the
explanatory variables [viz., log($\langle g \rangle$) and log($\langle
T \rangle$)] are not highly correlated
with each other. Thus we need to
explicitly test for the absence of significant collinearity, i.e.,
linear dependence of one explanatory variable on the other variables.
A commonly used indicator of collinearity is the variance inflation
factor ($VIF$)~\cite{Belsley1980}. 
When the variation of a specific explanatory variable
(referred to as a predictor) is largely explained by a linear
combination of the other predictors, VIF for that predictor is
correspondingly large. Complete absence of collinearity corresponds to
the case $VIF=1$ and the inflation is measured relative to this
reference value.
$VIF$ have been shown to correspond to the diagonal elements of the
inverse of the matrix of correlations between the
predictors~\cite{Belsley1980} and using this method we obtain
$VIF=1.28$ for both the macroeconomic factors considered by us.
As commonly collinearity is considered to be a cause for concern only
if VIF values are higher than 5, GDP per capita and Theil index 
can be reasonably treated as independent explanatory variables in our
analysis.
We have also investigated the possible dependence of the nature of the 
fluctuation distribution on other economic factors, such as the
foreign direct investment (FDI)
net inflow, but none of these appear to be independent of
the two factors considered above.

\begin{figure}
\begin{center}
\includegraphics[width=0.99\linewidth,clip]{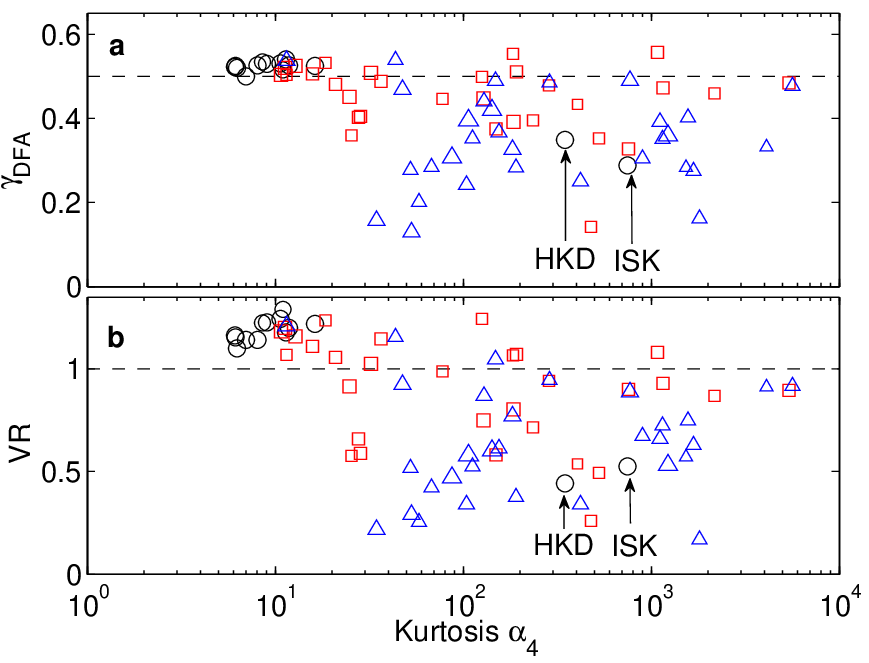} 
\end{center}
\caption{(color online).
{\bf Variation of 
(a) the long-range auto-correlation scaling exponent
$\gamma_{DFA}$ obtained using detrended fluctuation analysis of
the exchange rate time series, and
(b) the variance ratio ($VR$) of the exchange rate fluctuations
calculated using lag $l$ ($= 10$ days), 
with the kurtosis $\alpha_4$ of the normalized logarithmic return
distributions of different currencies.}
Different symbols are used to indicate currencies from developed
(black, circles), 
emerging (red, squares) and frontier (blue, triangles) economies, while symbol
size is proportional to
log($\langle g \rangle$) of the corresponding countries.
The broken lines in (a) and (b) indicate the values of $\gamma_{DFA}
(=0.5)$
and $VR (=1)$ corresponding to an uncorrelated random walk.
Currencies of developed economies that are outliers, viz., ISK and HKD
that have much higher kurtosis than others in the group, are
explicitly indicated.
}
\label{fig4}
\end{figure}

To investigate the reason for the strong relation between the kurtosis
of the return distribution for a currency and the corresponding
underlying macroeconomic factors, we need to delve deeper into the
nature of the dynamics of the exchange rate fluctuations. For this we
first look into the self-similar scaling behavior of the time-series
of exchange rate of a currency $P(t)$ using the detrended fluctuation
analysis (DFA) technique suitable for analyzing non-stationary processes 
with long-range memory~\cite{Peng}. Here, a time-series is de-trended
over different temporal windows of sizes $s$ using least-square
fitting with a linear function. The residual
fluctuations $F(s)$ of the resulting sequence, measured in terms of
the standard deviation, is seen to scale as $F(s) \sim s^{\gamma_{DFA}}$, 
where $\gamma_{DFA}$ is referred to as the DFA exponent. The numerical
value of this exponent (lying between 0 and 1) provides information
about the nature of the fractional Brownian motion undertaken by the
system. For $\gamma_{DFA} \simeq 1/2$, the process is said to be
equivalent to a random walk subject to white noise, while
$\gamma_{DFA} > 1/2$ ($<1/2$) implies that the time-series is
correlated (anti-correlated).
As seen from Fig.~\ref{fig4}~(a), the DFA exponents of currencies for 
most developed
economies - which also have the lowest kurtosis - are close to 0.5,
indicating that these currencies are following uncorrelated random
walk~\cite{Ausloos2000}. 
In contrast, currencies of the emerging and frontier economies,
possessing higher values of kurtosis, typically have $\gamma_{DFA} <
0.5$ indicating sub-diffusive dynamics.


To understand the reason for this sub-diffusive behavior we have
analyzed the exchange rate time-series using the 
variance ratio (VR) test. This technique, based on the ratio of
variance estimates for
the returns calculated using different temporal lags, is often used to
find how close a given time-series
is to a random walk~\cite{Lo1989}. For a sequence of log returns
$\{R_t\}$, the variance ratio for a lag $l$ is defined as:
\begin{equation}
VR (l)=
\frac{\sum_{k = l}^{\tau} (\sum_{t=k-l}^{k-1} R_t - l
\mu_R)^2}{\sigma_R^2 l (\tau- l +1)(1-[l/\tau])},
\end{equation}
where $\mu_R = \langle R_t \rangle$ and $\sigma_R^2 = \langle (R_t -
\mu_R)^2 \rangle$ are the mean and variance of the $\{R_t\}$ sequence.
An uncorrelated random walk is characterized by a VR value close to 1.
If $VR >1$, it indicates mean aversion in the time-series, i.e., the
variable has a tendency to follow a trend where successive changes are 
in the same direction. In contrast, $VR <1$ suggests a mean-reverting
series
where changes in a given direction are likely to be followed by
changes in the opposite direction preventing the system from moving
very far from its mean value.  Fig.~\ref{fig4}~(b) shows the VR values
for different currencies, calculated using lag $l = 10$ days, 
as a function of their kurtosis.
Consistent with the DFA results reported above, it is seen
that for currencies of developed economies the VR is close to 1,
indicating uncorrelated Brownian diffusion as the nature of their
exchange rate dynamics. However, for most frontier and a few emerging
economy currencies, the VR value is substantially smaller than 1,
implying that their trajectories have a mean-reverting nature.
As in Fig.~\ref{fig3}, we note that HKD and ISK appear as outliers in
Fig.~\ref{fig4} in that, although belonging to the group of
countries having high GDP per capita, they share the characteristics
shown by most emerging and frontier economies.

We can now understand the sub-diffusive nature of the dynamics of
these currencies as arising from the anti-correlated nature of their
successive fluctuations which prevents excursions far from the
average value. Thus, when we consider the time-series of all
currencies after normalizing their variance, the fluctuations of the 
emerging and frontier economy currencies mostly remain in the
neighborhood of the average value with rare, occasional
deviations that are very large compared to developed economy
currencies. This accounts for the much heavier tails of the return
distributions of the former and the corresponding high value of
kurtosis.
It is intriguing to consider whether the difference in the nature of
the movement of exchange rates of the
currencies could be possibly related to the role played by speculation
in the trading of these currencies~\cite{Brown2006}.
We also note that these results are in broad agreement with the fact
that efficient markets follow uncorrelated random walks and the notion
that the markets of developed economies are far more efficient than
those of emerging and frontier ones. 
A temporally resolved analysis of the nature of the distributions at
different periods shows strong disruption of the otherwise regular
pattern of systematic deviation during the severe crisis of 2008-09,
indicating its deep-rooted nature affecting the real
economy.

\subsection{Temporal evolution of system properties}
\begin{figure}[t]
\begin{center}
\includegraphics[width=0.99\linewidth,clip]{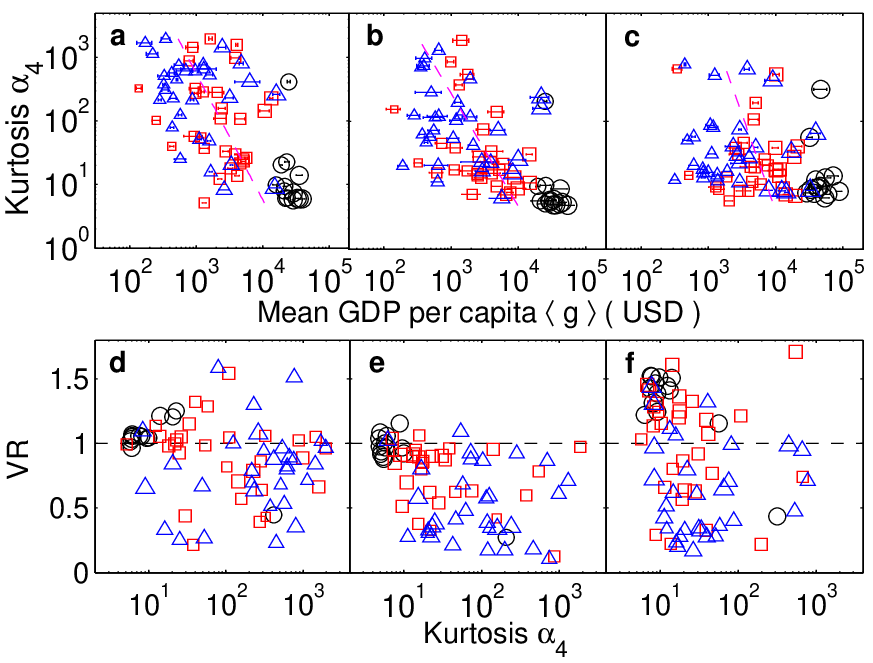} 
\end{center}
\caption{(color online).
{\bf Temporal evolution of the statistical
properties of
exchange rate fluctuation distributions of different currencies.}
The variation of (a-c) the kurtosis $\alpha_4$ of the distributions
with annual GDP per capita, $g$ (in USD) and that of (d-f) the
variance ratio (VR) of the different normalized fluctuations time series
with kurtosis $\alpha_4$, are shown for three different
periods, viz.,
Period I: Oct 23, 1995 - Apr 25, 2001 (a \& d), Period II: Apr 26, 2001 -
Oct 28, 2006 (b \& e) and Period III: Oct 29, 2006 - Apr 30, 2012 (c
\& f),
which divide the
duration under study into three equal, non-overlapping segments.
The GDP per capita of the different countries for each period are
obtained by averaging the annual values over the corresponding
periods.
The Pearson correlation coefficients between log($ \langle g \rangle$) and
log($\alpha_4$) for the three periods are $\rho_I = -0.60$ ($p$-value $=
10^{-8}$), $\rho_{II} = -0.57$ ($p$-value $= 10^{-8}$) and $\rho_{III} =
-0.28$ ($p$-value $= 10^{-2}$). For the first two periods, the best-fit
functional relation between the two is $\alpha_4 \sim 1/\langle g \rangle^2$,
while for the third period, the dependence of $\alpha_4$ on $\langle g
\rangle$ shows a strong deviation from the inverse square relation
seen in the other two periods.
Comparing the variance ratio values for the three different periods
show a higher degree of mean aversion in the third period.
Period III, during which the major economic crisis of 2008-09
occurred, is distinguished by large deviation from the trends seen in
the other two periods.
Different symbols are
used to indicate currencies from developed (black, circles), emerging
(red, squares) and frontier (blue, triangles) economies, while symbol size is
proportional to
log($\langle g \rangle$) of the corresponding countries.
}
\label{fig6}
\end{figure}

In the analysis presented above
we have considered the entire temporal duration
which our data-set spans. However, as the world economy underwent
significant changes during this period, most notably, the global
financial crisis of 2008, it is of interest to see how the
properties we investigate have evolved with time. For this purpose we
divide the data-set into three equal non-overlapping periods each
comprising 2011 days,
corresponding to Period I: Oct 23, 1995 - Apr 25, 2001, Period II:
Apr 26, 2001 - Oct 28, 2006 and Period III: Oct 29, 2006 - Apr 30, 2012.
Note that the last period corresponds to the crisis of the global
economy spanning 2007-2009.
For each of these, we carry out the same procedures as described earlier
in the context of the
the entire data-set.
As seen from Fig.~\ref{fig6}, the behavior in the first two intervals
appear to be quite similar in terms of the various properties that
have been measured, but large deviations are seen in the third
interval. This is apparent both for the relation between kurtosis and
mean GDP per capita [Fig.~\ref{fig6}~(a-c)], as well as that between
kurtosis and mean Theil index (figure not shown). The dependence
of the nature of the fluctuation distribution on the properties of the
underlying economy seem to have weakened in Period III.
For example, while there is significant strong negative correlation
between log($\langle g \rangle$) and log($\alpha_4$) for the first two intervals,
viz., $\rho = -0.60$ ($p=10^{-8}$) and $-0.57$ ($p=10^{-8}$),
respectively, it decreases to only $\rho = -0.28$ ($p=10^{-2}$) for
the third interval. Furthermore, the first two intervals show a $ 1/
\langle g \rangle^{2}$ dependence of the kurtosis $\alpha_4$, same as that seen for the
entire period that we have reported above.
However, this is not true for the last interval where
the best fit for the dependence of $\alpha_4$ on $\langle g\rangle$
shows a strong deviation from the behavior seen in other periods.
Similarly, we have found significant high
correlation between log($\langle T \rangle $) and log($\alpha_4$),
corresponding to Pearson coefficients $\rho =
0.50$ ($p = 10^{-6}$) and $\rho=0.46$ ($p = 10^{-5}$), respectively, for
the first two intervals. In contrast, for the third interval we
observe a relatively smaller correlation $\rho = 0.35$ ($p = 10^{-2}$).
In addition, the relation between the variance ratio and the kurtosis
of the returns [Fig.~\ref{fig6}~(d-f)], as well as that between the DFA
exponent and the kurtosis (figure not shown), are seen to be similar
in the first two intervals but very different in the third - in part
because the VR for the developed and some emerging economies have
adopted values $> 1$ (i.e., exhibiting mean aversion) in this last
interval, while earlier
they were close to 1 (i.e., similar to a random walk). While Periods I
and II had their share of economic booms and busts, it is
instructive to note that the 2008 crisis was severe enough to
disrupt systemic features that were otherwise maintained over time.

\section {Discussion and Conclusion} 
The work we report here underscores the importance of studying
economic systems, especially financial markets, for gaining an
understanding of the collective dynamics of heterogeneous complex
systems.
At the largest scale, such a system encompasses the entire
world where the relevant entities are the different national economies interacting
with each other through international trade and the foreign exchange
market. The far-from-equilibrium behavior of this highly heterogeneous
complex system has been investigated here by focusing on the fluctuations of
exchange rates of the respective currencies. Understanding the overall
features of this dynamics is crucially important in view of the human
and social cost associated with large-scale disruptions in the system,
as was seen during the recent 2008 world-wide economic crisis.

Our results suggest a putative invariant signature in the dynamics of
exchange rates, possibly the first such seen in 
macroeconomic phenomena. This is
in contrast to microeconomic systems like individual financial
markets where robust stylized facts such as the ``inverse cubic law"
has been established for some time.
The ``inverse square law" that we report here also has a fundamental
distinction in that distributions characterized by CCDF exponents $\alpha
\leq 2$ belong to the Levy-stable regime. By contrast, the
logarithmic return distributions of equities and indices of financial
markets that have exponent values
around $3$ are expected to converge to a Gaussian form at longer
time scales~\cite{Pan2008,Drozdz2003}. 
It suggests that extreme events corresponding to sudden
large changes in exchange rates, in particular for currencies
belonging to emerging and frontier economies, should be expected far more
often compared to 
other financial markets.
The ``inverse square law'' has recently been also reported
in at least one other market, viz., that of Bitcoins in the initial
period following its inception~\cite{Soumya2015}. We note that
agent-based modeling of markets suggest that such a
distribution can arise if market players are relatively homogeneous in
their risk propensity~\cite{Vikram2011,Feng2012}.

To conclude, the results of our study help in revealing a hidden
pattern indicative of relative invariance
in a highly heterogeneous complex system, viz., the FOREX market. The
robust empirical feature that we identify here is a power law
characterizing the heavy-tailed nature of the
fluctuation distributions
of exchange rates for different currencies.
The systematic deviation of individual currencies from the universal
form (the ``inverse square law''), quantified in terms of their kurtosis measuring the peakedness
of the return distributions, can be linked to metrics of the
economic performance and degree of
diversification of export products of the respective countries. 
By doing detrended fluctuation analysis, the distinct behavior of
currencies corresponding to developed, emerging and frontier markets
can be linked to the different scaling behaviors of the random walks
undertaken by these currencies. 
Our work shows how robust empirical regularities among the components of
a complex system can be uncovered even when the system is
characterized by a large number of heterogeneous
interacting elements exhibiting distinct local dynamics. 
Similar approaches may be used for identifying invariances in other biological and socio-economic systems.


\begin{acknowledgments}
We thank Anindya S. Chakrabarti, Tanmay Mitra and V. Sasidevan for helpful
suggestions. We gratefully acknowledge the assistance of Uday Kovur in the
preliminary stages of this work. This work was supported in part by IMSc
Econophysics (XII Plan) Project funded by the Department of Atomic Energy,
Government of India.
\end{acknowledgments}

\end{document}